\documentclass[acmsmall,screen]{acmart}

\pdfoutput=1  

\setcopyright{acmlicensed}
\copyrightyear{2026}
\acmYear{2026}
\acmDOI{XXXXXXX.XXXXXXX}

\acmVolume{XXX}
\acmNumber{YYY}
\acmArticle{111}
\acmMonth{1}

\newtheorem{theorem}{Theorem}[section]

\newtheorem{corollary}[theorem]{Corollary}

\theoremstyle{definition}
\newtheorem{definition}{Definition}[section]

\theoremstyle{remark}
\newtheorem{remark}{Remark}[section]

\newtheorem{theoremsection}{Theorem}[section]

\newtheorem{lemma}[theoremsection]{Lemma}

\def\QEDopen{{\setlength{\fboxsep}{0pt}\setlength{\fboxrule}{0.2pt}\fbox{\rule[0pt]{0pt}{1.3ex}\rule[0pt]{1.3ex}{0pt}}}}
\def\QED{\QEDopen}

\def\Q.E.D{\hfill\QED}

\begin{document} 

\title{Resolution of The Linear-Bounded Automata Question}%

 \date{\today}

\author{Tianrong Lin}
\affiliation{%
 \institution{Hakka University}
 \country{China}
}
\orcid{0000-0002-1187-2395}



\begin{abstract}
This paper resolves a famous and longstanding open question in automata theory, i.e., the {\it linear-bounded automata question} (or, for short, the LBA question), which can also be phrased succinctly in the language of computational complexity theory as $${\rm NSPACE}[n]\overset{?}{=}{\rm DSPACE}[n]. $$
In fact, we prove a more general result that $${\rm DSPACE}[S(n)]\subsetneqq {\rm NSPACE}[S(n)], $$where $S(n)\geq n$ is a space-constructible function. Our proof technique is based on diagonalization against deterministic $S(n)$ space-bounded Turing machines by means of a universal nondeterministic Turing machine, together with other novel and interesting new techniques developed in this paper. Our proof also implies the following consequences, which resolve some famous open questions in complexity theory:\\
\indent (1). ${\rm DSPACE}[n]\subsetneqq {\rm NSPACE}[n]$;\\
\indent (2). $L\subsetneqq NL$;\\
\indent (3). $L\subsetneqq P$;\\
\indent (4). There exists no deterministic Turing machine working in $O(\log n)$ space that decides the $st$-connectivity question (STCON). 
\end{abstract}



\received{17 August 2026}
\received[revised]{17 August 1960}
\received[accepted]{17 August 1960}

\setcopyright{rightsretained}
\acmYear{2026} \acmVolume{1} \acmNumber{1} \acmArticle{1} \acmMonth{1}\acmDOI{10.1145/3672614}

\maketitle
\vskip 0.3 cm
\section{Introduction}
\label{sec:introduction}

\subsection{Background}
\vskip 0.3 cm

The field of {\it automata theory} (see e.g., \cite{A3}), which was more or less influenced by Turing's abstraction \cite{Tur37} when viewing Turing machines as the most general kind of automata,\footnote{In \cite{Kle56}, Kleene's only reference to Turing is an explanation that Turing machines are not finite automata because of their unbounded tapes. Indeed, most (but not all) types of automata are special cases of the {\it Turing machine}, but our starting points are that Turing machines are a kind of automata whose memory, i.e., tape cells, is infinite, with its head being able to read, write, and move left or right; Furthermore, although linear-bounded automata are Turing machines, we still follow the terminologies in \cite{Myh60, Kur64} and call them linear-bounded $automata$ or, shortly, LBAs.} has witnessed two fundamental branches in the last eight decades, the first of which is the study of the notion of computability; see e.g., Rogers \cite{Rog67}, dealing with the main topics of what is computable by {\it Turing machines}. This kind of study, including other important questions related to Turing machines, has been extended to different types of automata such as {\it finite automata}, {\it pushdown automata}, {\it linear-bounded automata}, and so on; see e.g. Ginsburg \cite{Gin66}, Hopcroft et al. \cite{HU79}, and Kuroda \cite{Kur64}. It is worth noting that, in the context of {\it formal language theory}, a Turing machine is an automaton capable of enumerating some arbitrary subset of valid strings of an alphabet (see e.g., \cite{A4}).

As already observed by Stearns et al. \cite{SHL65}, an important goal of automata theory is a basic understanding of computational processes for various classes of problems. Thus, the theory of automata expanded beyond the notion of computability to include some measure of the difficulty of the computation and how that difficulty is related to the organization of the machine that performs the computation, which formed the second branch of automata theory known as {\it complexity theory}; see e.g., \cite{SHL65}.

Perhaps the most basic and fundamental measures of difficulty in {\it complexity theory} that appear particularly important are time and space (memory). Clearly, the study of space complexity classes ${\rm DSPACE}[S(n)]$, ${\rm NSPACE}[S(n)]$, and ${\rm coNSPACE}[S(n)]$ was made possible by the basic measure of space, which also enabled the theory of ${\rm PSPACE}$-completeness; see e.g., \cite{GJ79,AB09,Pap94,Sip13}. Today, these notions are important building blocks in complexity theory.

As is well-known, nondeterminism is another important notion in {\it complexity theory}. The question of whether nondeterminism adds power to a space-bounded Turing machine was apparently first posed by Kuroda \cite{Kur64}, who asked whether deterministic and nondeterministic {\it linear bounded automata} are equivalent. Indeed, in the 1960s, in his seminal paper \cite{Kur64}, Kuroda, who showed that the class of languages accepted by {\it nondeterministic linear-bounded automata} (i.e., the complexity class ${\rm NSPACE}[n]$) is precisely the context-sensitive languages, stated two research challenges, which subsequently became famously known as the LBA questions (see e.g., \cite{HH73,HH74} for a discussion of the history and importance of these questions), the first of which is whether the class of languages accepted by nondeterministic linear-bounded automata is equal to the class of languages accepted by deterministic linear-bounded automata, i.e., the following:
\begin{center}
\fbox{\parbox{\textwidth}{{\bf Question 1:}  Are the complexity classes ${\rm NSPACE}[n]$ and ${\rm DSPACE}[n]$ the same? }}
\end{center}

If ${\rm NSPACE}[n]$ and ${\rm DSPACE}[n]$ are the same, then the family of context-sensitive languages is closed under complement. On the other hand, it still could happen that $${\rm DSPACE}[n]\neq {\rm NSPACE}[n]. $$ But the family of context-sensitive languages is closed under complement. Thus we have the second LBA question, i.e., the following:

\begin{center}
\fbox{\parbox{\textwidth}{{\bf Question 2:}  Are the complexity classes ${\rm NSPACE}[n]$ and ${\rm coNSPACE}[n]$ the same? }}
\end{center}

The above two questions stated respectively can be phrased succinctly in the language of computational complexity theory as (see e.g., \cite{A2}):
$$\aligned
   {\rm NSPACE}[n]\overset{?}{=}&{\rm DSPACE}[n];\\
   {\rm NSPACE}[n]\overset{?}{=}&{\rm coNSPACE}[n].
\endaligned$$

Both of these questions are basically problems about the minimal amount of memory needed to perform a computation. As said in \cite{HH73}, such questions, in general, are quite difficult. Although the second LBA question (i.e., Question 2) has an affirmative answer --- implied by the famous Immerman-S\'{e}nizergues theorem (see \cite{Imm88,Sze88}) --- proved $20$ years after the question was raised, the first LBA question (i.e., Question 1) still remains open. It is worth noting that Kuroda \cite{Kur64} showed that the class of languages accepted by {\it nondeterministic linear-bounded automata} (i.e., the complexity class ${\rm NSPACE}[n]$) is precisely the context-sensitive languages, so the first LBA question is equivalent to asking whether the whole of the context-sensitive languages can be accepted by deterministic linear-bounded automata; see e.g., \cite{A2}. In fact, Question $1$ has become one of the oldest open questions in {\it automata theory} and {\it complexity theory}; see e.g., \cite{Sav70,RCH91}. At the same time, our inability to answer this question indicates that we have not yet understood the nature of nondeterministic computation (see \cite{HH73}). In what follows, as the title indicates, we will call the first LBA question, which is still open, the LBA question.

Interestingly, Savitch's algorithm \cite{Sav70} gives us an initial insight into the LBA question: nondeterministic space-bounded Turing machines can be simulated efficiently by deterministic space-bounded Turing machines, with only a quadratic loss in space usage. That is, $${\rm NSPACE}[S(n)]\subseteq{\rm DSPACE}[S(n)^2] $$ for $S(n)\geq\log n$,\footnote {Throughout this paper, $\log n$ stands for $\log_2 n$.} where $S(n)$ is a {\it space-constructible function}. As can be seen, the technique used in the proof of Savitch's algorithm \cite{Sav70} is an interesting application of divide-and-conquer (see e.g., p. 369 in \cite{AHU74}).

In addition, although most of the interesting questions related to the power of nondeterminism remain open, and we still do not know whether nondeterministic space is equal to deterministic space, we believe that nondeterministic space is more powerful than deterministic space. Indeed, Cook et al. \cite{CR80} have already given some evidence that ${\rm NSPACE} [\log n]$ is more powerful than ${\rm DSPACE} [\log n]$ by showing that the limited model of $\log$-space Turing machine (JAG machines) cannot recognize the Threadable Mazes set, where the Threadable Mazes set introduced by Savitch \cite{Sav70} is $\log$-space complete for ${\rm NSPACE}[\log n]$. Nonetheless, we still lack proof of the exact relation between the complexity class ${\rm DSPACE}[\log n]$ (i.e., $L$) and the complexity class ${\rm NSPACE}[\log n]$ (i.e., $NL$), so we have the following:

\begin{center}
\fbox{\parbox{\textwidth}{{\bf Question 3:}  Are the complexity classes $\mathit{L}$ and $\mathit{NL}$ the same?  }}
\end{center}

The directed $st$-connectivity question, denoted STCON, is one of the most widely studied questions in theoretical computer science and is simple to state: Given a directed graph $G$ together with vertices $s$ and $t$, the $st$-connectivity question is to determine if there is a directed path from $s$ to $t$. Interestingly, STCON plays an important role in complexity theory as it is complete for the complexity class {\it nondeterministic logspace} $NL$ under log-space reductions; see e.g., \cite{AB09,Mic92,Pap94,Sip13}. Currently, the best-known space upper bound is $O(\log ^2 n)$ using Savitch's algorithm \cite{Sav70}. In a breakthrough result for the undirected $st$-connectivity problem, Reingold \cite{Rei08} showed, using the Zig-Zag graph product \cite{RVW02}, that undirected $st$-connectivity can be solved in $O(\log n)$ space. This renews enthusiasm for improving Savitch's bound for STCON, since one obvious directions is to extend Reingold's algorithm to the directed case. However, it is considered that proving any nontrivial $\Omega(\log n)$ space lower bound on a general Turing machine is beyond the reach of current techniques. So, we have the following open question:

\begin{center}
\fbox{\parbox{\textwidth}{{\bf Question 4:}  Is there a deterministic $O(\log n)$ space-bounded Turing machine deciding the STCON?  }}
\end{center}

\vskip 0.3 cm
\subsection{Main Results}
\vskip 0.3 cm

In this paper, we will work with the techniques of diagonalization against deterministic space-bounded Turing machines with a universal nondeterministic Turing machine to resolve the LBA question. The diagonalization technique \cite{Tur37, HS65} is a standard method to prove lower bounds on uniform computing models (i.e., the Turing machine model). These techniques work well in deterministic time and space measures; see e.g., \cite{For00, FS07}. For nondeterministic time, we can still do a simulation but can no longer negate the answer directly. In this case, we apply the {\em lazy diagonalization} to show the nondeterministic time hierarchy theorem; see e.g., \cite{AB09,SFM78,Zak83}. For more information about the diagonalization, see e.g., \cite{For00, FS07} or Turing's original article \cite{Tur37}). We remark that the techniques of diagonalization against deterministic Turing machines by a universal nondeterministic Turing machine will be further used in the author's other works, such as \cite{Lin21a,Lin21b}, since both of them handle the topic of determinism versus nondeterminism. Before getting into the main contributions, let us make some digressions. The inspiration of diagonalization against deterministic Turing machines by a universal nondeterministic Turing machine in this paper originated from two concurrent matters: At that moment,\footnote{Around an afternoon in early October 2021.} the author was reading the proof of the {\it Space Hierarchy Theorem for Deterministic Turing Machines} in \cite{AHU74} (see Theorem 11.1 in standard textbook \cite{AHU74}, p. 408--410); and simultaneously, the author was considering the LBA question. Then, a very natural idea appears: If one uses a universal nondeterministic Turing machine to diagonalize against all deterministic $S(n)$ space-bounded Turing machines, is it possible to obtain a language in ${\rm NSPACE}[S(n)]$? If so, setting $S(n)=n$ will yield a negative answer to the LBA question. In fact, all conclusions in this paper are the direct or indirect products of this inspiration.

Indeed, the following is our first main result:
\begin{theorem}
\label{theorem1}
  Let $S(n)\geq n$ be a space-constructible function. Then there is a language $L_d$ such that $$L_d\in {\rm NSPACE}[S(n)]$$ but $$L_d\not\in {\rm DSPACE}[S(n)]. $$ That is, $${\rm DSPACE}[S(n)]\subsetneqq {\rm NSPACE}[S(n)]. $$
\end{theorem}

Focusing our attention on the LBA question; only the limited case, i.e., the Turing machines are limited to the one-way case but not the two-way, was shown in \cite{HU67} that the nondeterministic models are more powerful than deterministic models. However, for two-way Turing machines of all other complexity classes, it is an open question as to whether or not the deterministic and nondeterministic models are equivalent. In fact, set $S(n)=n$; then the above Theorem \ref{theorem1} immediately yields a proof of a negative answer to the LBA question:

\begin{theorem}
\label{theorem2}
There exists a language $L_d$ such that $$L_d\in {\rm NSPACE}[n]$$ but $$L_d\not\in {\rm DSPACE}[n]. $$ That is, $${\rm DSPACE}[n]\subsetneqq {\rm NSPACE}[n]. $$
\end{theorem}

For the complexity classes $L$, $NL$, and the deterministic polynomial-time class $P$, and so on, there is a well-known tower of inclusions; see e.g., \cite{AB09,Pap94,Sip13}: $$\mathit{L}\subseteq\mathit{NL}\subseteq\mathit{P}\subseteq\mathit{NP}\subseteq\mathit{ PSPACE}. $$

It is unknown whether $\mathit{L}\subsetneqq \mathit{NL}$ and $\mathit{L}\subsetneqq \mathit{P}$; see e.g., \cite{A1}. With the above Theorem \ref{theorem2} at hand, we are able to resolve the famous open question $L\overset{?}{=}NL$ (see e.g.,\cite{AB09,Pap94,Sip13}) by showing the following:

\begin{theorem}
\label{theorem3}
$\mathit{L}\subsetneqq \mathit{NL}$. That is, $$ \mathit{L}\neq \mathit{NL}. $$
\end{theorem}
\noindent It immediately follows from the above that

\begin{corollary}
\label{corollary1}
$\mathit{L}\neq \mathit{P}$. That is, $$\mathit{L}\subsetneqq \mathit{P}. $$
\end{corollary}

However, currently it is unknown whether $NL=P$, and to the best of our knowledge, it is also an interesting open question in theoretical computer science.

Furthermore, the question of $\mathit{L}\overset{?}{=}\mathit{NL}$ and STCON are closely correlated, since STCON is {\it $\log$-space complete} for $\mathit{NL}$ (see e.g., \cite{AB09,Pap94,Sip13}). Thus, if STCON can be decided in space $O(\log n)$ deterministically, then $\mathit{L}=\mathit{NL}$ will follow, and vice versa. It is clear that the above Theorem \ref{theorem3}, together with some basic knowledge, can show the following:

\begin{theorem}
\label{theorem4}
There exists no deterministic $O(\log n)$ space-bounded Turing machine deciding the STCON.
\end{theorem}

\vskip 0.3 cm
\subsection{Overview}
\vskip 0.3 cm

The remainder of this paper is organized as follows: For the convenience of the reader, in the next section, we will review some notation and notions that are closely associated with our introductions appearing in Section \ref{sec:introduction}. In Section \ref{sec:diagonalization} and Section \ref{sec:proof_of_ld_in_nspacesn} we will prove our more general results, which lead to our main results. The proofs of Theorem \ref{theorem1} and Theorem \ref{theorem2} are given in Section \ref{sec:proofs_theorem1_theorem2}. In Section \ref{sec:L_NL}, we resolve the famous open question of $L$ vs. $NL$ by giving the proof of Theorem \ref{theorem3}. Section \ref{sec:proofs_theorem4} is dedicated to solving STCON by showing Theorem \ref{theorem4}. Finally, a brief conclusion is drawn, and some open questions are introduced in the last section.

\vskip 0.3 cm
\section{Preliminaries}
\label{sec:preliminaries}
\vskip 0.3 cm

In this section, we describe some notions and notation needed in the following context.

Let $S$ be a finite set; then $|S|$ denotes the cardinality of $S$.

Let $\mathbb{N}$ denote the natural numbers: $$\{0,1,2,3,\cdots\}, $$where $+\infty\not\in\mathbb{N}$. Further, $\mathbb{N}_1$ denotes the set of $$\mathbb{N}\setminus\{0\}, $$i.e., the set of positive integers. 

For convenience, we also denote the set of positive integers that are greater than or equal to $3$ by $\mathbb{N}_3$, i.e., $$\mathbb{N}_3\overset{\rm def}{=}\mathbb{N}\setminus\{0,1,2\}. $$

Throughout the paper, the computational modes used are {\it Turing machines} and {\it linear-bounded automata}. We follow the definition of a Turing machine given in the standard textbook \cite{AHU74} and redefine the {\it linear-bounded automata} according to the notion of {\it space-bounded Turing machine}.

\begin{definition}[$k$-tape deterministic Turing machine, \cite{AHU74}]
\label{definition2.1}
 A $k$-tape deterministic Turing machine (shortly, DTM) $M$ is a seven-tuple $(Q,T,I,\delta,\mathrm{b},q_0,q_f)$ where:
 \begin{enumerate}
 \item {$Q$ is the set of states.}
 \item {$T$ is the set of tape symbols.}
 \item {$I$ is the set of input symbols; $I\subseteq T$.}
 \item {$\mathrm{b}\in T-I$, is the blank.}
 \item {$q_0$ is the initial state.}
 \item {$q_f$ is the final (or accepting) state.}
 \item {$\delta$ is the next-move function, which maps a subset of $Q\times T^k$ to $$Q\times(T\times\{L,R,S\})^k. $$
  That is, for some $(k+1)$-tuples consisting of a state and $k$ tape symbols, it gives a new state and $k$ pairs, each pair consisting of a new tape symbol and a direction for the tape head. Suppose $$\delta(q,a_1,a_2,\cdots,a_k)=(q',(a'_1,d_1),(a'_2,d_2),\cdots,(a'_k,d_k)), $$ and the deterministic Turing machine is in state $q$ with the $i$th tape head scanning tape symbol $a_i$ for $1\leq i\leq k$. Then in one move the deterministic Turing machine enters state $q'$, changes symbol $a_i$ to $a'_i$, and moves the $i^{th}$ tape head in the direction $d_i$ for $1\leq i\leq k$.}
  \end{enumerate}
 \end{definition}

The notion of a nondeterministic Turing machine is similar to that of a deterministic Turing machine, except that the next-move function $\delta$ is a mapping from $Q\times T^k$ to subsets of $Q\times(T\times\{L,R,S\})^k$, stated as follows:

\begin{definition}[$k$-tape nondeterministic Turing machine, \cite{AHU74}]
\label{definition2.2}
A $k$-tape nondeterministic Turing machine (shortly, NTM) $M$ is a seven-tuple $(Q,T,I,\delta,\mathrm{b},q_0,q_f)$ where all components have the same meaning as for the ordinary deterministic Turing machine, except that here the next-move function $\delta$ is a mapping from $Q\times T^k$ to subsets of $Q\times(T\times\{L,R,S\})^k$.
\end{definition}

The {\it language accepted by} the Turing machine $M$, denoted as $L(M)$, is the set of words $w$ in $I^*$ such that $M$ enters an accepting state. If for every input word of length $n$, deterministic (resp. nondeterministic) Turing machine $M$ scans at most $S(n)$ tape cells on any storage tape (work-tape), then $M$ is said to be a deterministic (resp. nondeterministic) $S(n)$ {\it space-bounded Turing machine}, or {\it of space complexity} $S(n)$. The language recognized by $M$ (i.e., $L(M)$) is also said to be of space complexity $S(n)$.

We should note that if a Turing machine can rewrite on its input tape, then it is an {\it on-line} Turing machine (as in Definition \ref{definition2.1} and Definition \ref{definition2.2}). Otherwise, it is an {\it off-line} Turing machine. An online Turing machine $M$ is of space complexity $S(n)$ if for each input of length $n$, $M$ uses at most $S(n)$ tape cells of its storage tape (including input tape). For an {\it off-line} Turing machine $M$, it has a read-only input tape and $k$ semi-infinite storage tapes (work tapes). If for every input word $w$ of length $n$, $M$ scans at most $S(n)$ cells on any storage tape, then $M$ is said to be an $S(n)$ space-bounded off-line Turing machine, or of space complexity $S(n)$. Then, a language is of space complexity $S(n)$ for some machine model if it is defined by a Turing machine of that model, which is of space complexity $S(n)$.

Also note that {\it off-line} Turing machines enable us to consider space bounds of less than linear growth. For example, in Section \ref{sec:L_NL}, we will deal with {\it off-line} Turing machines, since their space bounds are less than linear growth, namely $\log n$. It is worth noting that if a Turing machine could rewrite on its input tape, then the length of the input would have to be included in calculating the space bound. Thus no space bound could be less than linear (see e.g., \cite{HU79}), i.e., its space bounds $\geq n$.

The notation ${\rm DSPACE}[S(n)]$ and ${\rm NSPACE}[S(n)]$ denote the class of languages accepted by deterministic $S(n)$ space-bounded Turing machines and nondeterministic $S(n)$ space-bounded Turing machines, respectively. In particular, $L$ is the class ${\rm DSPACE}[\log n]$ and $NL$ is the class ${\rm NSPACE}[\log n]$, i.e., the class of languages accepted by deterministic and nondeterministic $\log n$ space-bounded Turing machines, respectively.

The original notion of a deterministic (nondeterministic) {\it linear-bounded automaton} is a deterministic (nondeterministic) single-tape (i.e., {\it on-line}) Turing machine whose read/write head never leaves those cells on which the input was placed; see \cite{Myh60,Lan63,Kur64}. The equivalent definition of {\it linear-bounded automaton} is as follows:

\begin{definition}
\label{definition2.3}
Formally, a deterministic (nondeterministic) {\it linear-bounded automaton} is an {\it on-line} deterministic (nondeterministic) $S(n)=n$ space-bounded Turing machine.
\end{definition}

The following two lemmas are convenient tools needed in the proof of the main results, whose proof can be found on page $372$ of \cite{AHU74}:

\begin{lemma}(Corollary 3, \cite{AHU74}, p. 372)
\label{lemma1}
If $L$ is accepted by a $k$-tape deterministic Turing machine of space complexity $S(n)$, then $L$ is accepted by a single-tape deterministic Turing machine of space complexity $S(n)$.\Q.E.D
\end{lemma}

\begin{lemma}(Corollary 2, \cite{AHU74}, p. 372)
\label{lemma2}
If $L$ is accepted by a $k$-tape nondeterministic Turing machine of space complexity $S(n)$, then $L$ is accepted by a single-tape nondeterministic Turing machine of space complexity $S(n)$.\Q.E.D
\end{lemma}

By Lemma \ref{lemma2}, we know that if a language $L$ is accepted by an {\it on-line} $k$-tape deterministic Turing machine of space complexity $S(n)\geq n$, then $L$ is accepted by an {\it on-line} single-tape deterministic $S(n)$ space-bounded Turing machine. Thus we can restrict ourselves to ({\it on-line}) single-tape deterministic Turing machines when studying the LBA question.

For a complexity class $\mathcal{C}$, its complement is denoted by ${\rm co}\mathcal{C}$ (see e.g., \cite{Pap94}), i.e., $${\rm co}\mathcal{C}=\{\overline{L}\,:\,L\in\mathcal{C}\}, $$where $L$ is a decision problem, and $\overline{L}$ is the complement of $L$. Note that the complement of a decision problem $L$ is defined as the decision problem whose answer is ``{\it yes}" whenever the input is a ``{\em no}" input of $L$,\footnote{In this paragraph, $L$ is not the complexity class ${\rm DSPACE}[\log n]$ but a language accepted by some Turing machine.} and vice versa.

The following famous result, which resolved the second LBA question, says that the working space for both online and offline nondeterministic Turing machines is closed under complement:

\begin{lemma}(Immerman-S\'{e}nizergues Theorem, \cite{Imm88,Sze88})\label{lemma3}
Let $S(n)\geq\log n$ be a space-constructible function. Then the nondeterministic space of $S(n)$ is closed under complement. That is $${\rm NSPACE}[S(n)]={\rm coNSPACE}[S(n)]. $$
\end{lemma}

Other background information and notions will be given along the way in proving our main results stated in Section \ref{sec:introduction}.

\vskip 0.3 cm
\section{Diagonalization against Deterministic $S(n)$ Space-Bounded Turing Machines}
\label{sec:diagonalization}
\vskip 0.3 cm

To establish Theorem \ref{theorem1}, we need to {\it enumerate} the deterministic Turing machines, that is, assign an ordering to DTMs so that for each nonnegative integer $i$, there is a unique DTM associated with $i$.

By Lemma \ref{lemma1}, we can restrict ourselves to single-tape deterministic Turing machines. So, in the following context, by DTMs we mean single-tape DTMs. Thus, these DTMs are {\it on-line} Turing machines.

We will use the method presented in \cite{AHU74}, p. 407, to encode a single-tape deterministic Turing machine into an integer. Moreover, without loss of generality, we can make the following assumptions about the representation of a single-tape deterministic Turing machine with input alphabet $\{0,1\}$ because that will be all we need:
\begin{enumerate}
\item {The states are named
$$
q_1,q_2,\cdots,q_s
$$
for some $s$, with $q_1$ the initial state and $q_s$ the accepting state.}
\item {The input alphabet is $\{0,1\}$.}
\item {The tape alphabet is
$$
\{X_1,X_2,\cdots,X_t\}
$$
     for some $t$, where $X_1=\mathrm{b}$, $X_2=0$, and $X_3=1$.\footnote{ By this condition, it is clear that $t$, the number of tape symbols of a single-tape Turing machine, is a fixed positive integer in $\mathbb{N}_3$. And, different single-tape Turing machine may be with different $t\in\mathbb{N}_3$.}}
\item {The next-move function $\delta$ is a list of quintuples of the form
$$
         (q_i,X_j,q_k,X_l,D_m),
$$
          meaning that
$$
     \delta(q_i,X_j)=(q_k,X_l,D_m),
$$
and $D_m$ is the direction, $L$, $R$, or $S$, if $m=1,2$, or $3$, respectively. We assume this quintuple is encoded by the string
$$
     10^i10^j10^k10^l10^m1.
$$}
\item {The deterministic Turing machine itself is encoded by concatenating in any order the codes for each of the quintuples in its next-move function. Additional $1$'s may be prefixed to the string if desired. The result will be some string of $0$'s and $1$'s, beginning with $1$, which we can interpret as an integer.}
\end{enumerate}

By this method, any integer that cannot be decoded is assumed to represent the trivial Turing machine with an empty next-move function by this encoding. It is obvious that such a representation of a deterministic Turing machine defines a function that is surjective $$e:\mathbb{N}_1\rightarrow \mathcal{T}, $$where $\mathcal{T}$ is the set of all single-tape deterministic Turing machines. Hence, $e$ is an enumeration of the set of all single-tape deterministic Turing machines. In addition, we denote the $i^{\rm th}$ Turing machine in enumeration $e$ as $e(i)$. It should be noted that every DTM will appear infinitely often in the enumeration, since given a DTM, we may prefix $1$'s at will to find larger and larger integers representing the same set of quintuples.

We can now design a four-tape NTM $M$ that treats its input string $x_i$ both as an encoding of a DTM $M_i$ and also as the input to $M_i$. One of the capabilities possessed by $M$ is the ability to simulate a Turing machine, given its specification. We shall have $M$ determine whether the Turing machine $M_i$ accepts the input $x_i$ without using more than $S(|x_i|)$ tape cells for some space-constructible function $S$. If $M_i$ accepts $x_i$ in space $S(|x_i|)$, then $M$ does not. Otherwise, $M$ accepts $x_i$. Thus, for all $i$, either $M$ disagrees with the behavior of the $i^{\rm th}$ DTM on that input $x_i$, which is the binary representation of $i$, or the $i^{\rm th}$ DTM uses more than $S(|x_i|)$ tape cells on input $x_i$. We first show the following important result:

\begin{theorem}
\label{theorem3.1}
Let $S(n)\geq n$ be a space-constructible function. Then, there exists a language $L_d$ accepted by a nondeterministic Turing machine $M$ running within space 
$$
(1+\lceil\log t\rceil)S(n)\,\,\,\text{ for all $t\in\mathbb{N}_3$},
$$ 
but by no deterministic $S(n)$ space-bounded Turing machines. That is, $$L_d\notin{\rm DSPACE}[S(n)].$$
\end{theorem}

\begin{proof}
Let $M$ be a four-tape NTM that operates as follows on an input string $x$ of length $n$.
\begin{enumerate}
\item{ By using 
$$
O(\log |x|)
$$
space, $M$ decodes the single-tape deterministic Turing machine encoded by $x$. If $x$ is not the encoding of some single-tape deterministic Turing machine, then GOTO $6$. Otherwise, $M$ decodes the deterministic Turing machine encoded by $x$ and determines $t$, the number of tape symbols used by this deterministic Turing machine, and $s$, its number of states. The third tape of $M$ can be used as ``scratch" memory to calculate $t$. }
\item {$M$ marks off $(1+\lceil\log t\rceil)S(n)$ cells on each work tape. After doing so, if any tape head of $M$ attempts to move off the marked cells, $M$ halts without accepting.}
\item {Let $M_i$ be the deterministic Turing machine encoded by $x$.  $M$ then lays off on its second tape $S(n)$ blocks of $\lceil\log t\rceil$ cells each, the blocks being separated by single cells holding a marker $\#$, i.e., there are $(1+\lceil\log t\rceil)S(n)$ cells in all. Each tape symbol occurring in a cell of $M_i$'s tape will be encoded as a binary number in the corresponding block of the second tape of $M$. Initially, $M$ places its input, in binary-coded form, in the blocks of tape $2$, filling the unused blocks with the code for the blank.}
\item {On tape $3$, $M$ sets up a block of $\lceil\log s\rceil+\lceil\log S(n)\rceil+\lceil\log t\rceil S(n)$ cells, initialized to all $0$'s, provided again that this number of cells does not exceed $(1+\lceil\log t\rceil)S(n)$, which is possible, because $$\aligned
            \frac{\lceil\log s\rceil+\lceil\log S(n)\rceil+\lceil\log t\rceil S(n)}{(1+\lceil\log t\rceil S(n)}\rightarrow& \quad\frac{\lceil\log t\rceil}{1+\lceil\log t\rceil}\\
            <&\quad 1
             \endaligned$$ as $$ n\rightarrow +\infty
            $$
            and since there are arbitrarily long binary strings representing the same Turing machine $M_i$.
            Tape $3$ is used as a counter to count up to $sS(n)t^{S(n)}$.\footnote{If a single-tape deterministic/nondeterministic Turing machine $N$ (with $t$ tape symbols) is of constructible space complexity $S(n)$, then there is a number of distinct configurations that $N$ needs to enter when started with an input of length $n$, and this number is at most
             $$
            |Q|\times S(n)\times t^{S(n)},
            $$ 
            where $Q$ is the set of states of $N$.}
            }
\item {$M$ simulates $M_i$, using tape $1$ (its input tape) to determine the moves of $M_i$ and using tape $2$ to simulate the tape of $M_i$. The moves of $M_i$ are counted in binary in the block of tape $3$, and tape $4$ is used to hold the state of $M_i$. If $M_i$ accepts, then $M$ halts without accepting. $M$ accepts if $M_i$ halts without accepting. Moreover, $M$ will halt without accepting if the simulation by $M$ attempts to use more than the allotted cells on tape $2$, or if the counter on tape $3$ overflows, i.e., the number of moves made by $M_i$ exceeds $sS(n)t^{S(n)}$.}\label{item5}
\item{Since $x$ is not the encoding of some single-tape deterministic Turing machine, $M$ marks off 
$$
S(n)
$$
cells on each work tape. After doing so, if any tape head of $M$ attempts to move off the marked cells, $M$ halts without accepting. Then, on tape $3$, $M$ sets up a block of 
$$
S(n)
$$
cells, initialized to all $0$'s. Tape $3$ is used as a counter to count up to 
$$
2^{S(n)}.
$$

By using its nondeterministic choices, $M$ moves according to the path described by $x$. The moves of $M$ are counted in binary in the block of tape $3$. $M$ rejects if the number of moves made by $M$ exceeds 
$$
2^{S(n)}
$$
or $M$ reaches a reject state before the counter of tape $3$ reaches 
$$
2^{S(n)}.
$$
Otherwise, $M$ accepts. }
\end{enumerate}

The nondeterministic Turing machine $M$ described above is of space complexity
$$
(1+\lceil\log t\rceil)S(n)\quad\text{for all $t\in\mathbb{N}_3$},
$$
since $M_i$ has $t$ tape symbols with $t\in\mathbb{N}_3$ (see {\it footnote} $5$). It is worth noting that $M$ running within space
$$
(1+\lceil\log t\rceil)S(n)\quad\text{for all $t\in\mathbb{N}_3$}
$$
includes the most extreme case, i.e., the case that for each $t\in\mathbb{N}_3$, there exists a single-tape deterministic $S(n)$ space-bounded Turing machine with $t$ tape symbols. 

Then, by Lemma \ref{lemma2}, $M$ is equivalent to a single-tape NTM of space complexity 
$$
O((1+\lceil\log t\rceil)S(n))\quad\text{for all $t\in\mathbb{N}_3$},
$$
and it of course accepts some language $L_d$.

Suppose now $L_d$ were accepted by some deterministic $S(n)$ space-bounded Turing machine $M_j$. By Lemma \ref{lemma1}, we may assume that $M_j$ is a single-tape Turing machine. Let $M_j$ have $s$ states and $t$ tape symbols; we can find a string $w$ of length $n$ representing $M_j$ such that
$$
\frac{\lceil\log s\rceil+\lceil\log S(n)\rceil+\lceil\log t\rceil S(n)} {(1+\lceil\log t\rceil)S(n)}<1.
$$
This is also possible because
$$\aligned
\lim_{n\rightarrow+\infty}\frac{\lceil\log s\rceil+\lceil\log S(n)\rceil+\lceil\log t\rceil S(n)}{(1+\lceil\log t\rceil)S(n)}=&\frac{\lceil\log t\rceil}{1+\lceil\log t\rceil}\\
<&1.
\endaligned
$$
So, $M$ has enough spaces to simulate $M_j$ and accepts $w$ if and only if $M_j$ does not accept it. But we assumed that $M_j$ accepted $L_d$, i.e., $M_j$ agreed with $M$ on all inputs. We thus conclude that $M_j$ does not exist, i.e., $L_d$ is not accepted by any deterministic $S(n)$ space-bounded Turing machine.

Now, for convenience, if we denote by 
$$
{\rm NSPACE}\big[(1+\lceil\log t\rceil)S(n)\,\,\,\text{for all $t\in\mathbb{N}_3$}\big]
$$
the set of languages accepted by nondeterministic Turing machines that run within space
$$
(1+\lceil\log t\rceil)S(n)\,\,\,\text{for all $t\in\mathbb{N}_3$},
$$
we thus can conclude, by the above arguments, that
$$
{\rm DSPACE}[S(n)]\subsetneqq {\rm NSPACE}\big[(1+\lceil\log t\rceil)S(n)\,\,\,\text{for all $t\in\mathbb{N}_3$}\big].
$$
This completes the proof.
\end{proof}

\vskip 0.3 cm
\begin{remark}
In fact, we can design our universal nondeterministic Turing machine $M$ to be more complicated. For example, since we can also encode any nondeterministic Turing machine into a binary string (see e.g., {\rm \cite{Lin21b}}), the input to $M$ can be classified into three types: If the input is a deterministic Turing machine, then $M$ does the work specified in the proof of {\rm Theorem \ref{theorem3.1}}. If the input is a nondeterministic Turing machine $N$, then $M$ determines $t$, the number of tape symbols used by $N$, and $s$, its number of states, and then $M$ marks off $$(1+\lceil\log t\rceil)S(n)$$ cells on each tape and sets the counter of tape $3$ to count up to (see footnote $6$) $$sS(n)t^{S(n)}. $$

Then $M$ simulates $N$ nondeterministically, using tape $1$ (its input tape) to determine the moves of $N$ and using tape $2$ to simulate the tape of $N$. The moves of $M$ are counted in binary in the block of tape $3$, and tape $4$ is used to hold the state of $N$. $M$ accepts the input if and only if $N$ accepts. However, if the input cannot be decoded to some deterministic Turing machine or nondeterministic Turing machine, $M$ rejects the input. Note that such a design does not change $M$'s space complexity.
\end{remark}

\vskip 0.3 cm
\section{Proof of $L_d\in\text{NSPACE}[S(n)]$}
\label{sec:proof_of_ld_in_nspacesn}
\vskip 0.3 cm

Since the nondeterministic Turing machine $M$ constructed above, which accepts the language $L_d$, runs within space $(1+\lceil\log t\rceil)S(n)$ for all $t\in\mathbb{N}_3$. For the goal of this section, we define the following family of languages $\{L_d^i\}_{i\in\mathbb{N}_3}$:
$$\aligned
L_d^i\overset{{\rm def}}{=}&\mbox{ the language accepted by $M$ running within $(1+\lceil\log i\rceil)S(n)$ space for fixed $i\in\mathbb{N}_3$.}\\
         &\mbox{ That is, $M$ first marks off $(1+\lceil\log i\rceil)S(n)$ cells on each work tape before acting on item (1) }\\
          &\mbox{ in the proof of Theorem \ref{theorem3}. Then, $M$ is forced to halt if any work-tape head of}\\
          &\mbox{  $M$ attempts to move off the marked $(1+\lceil\log i\rceil)S(n)$ cells without accepting,}
\endaligned$$
which means that when $M$ is acting on item (2) within the $(1+\lceil\log i\rceil)S(n)$ cells previously marked, $M$ will reject the input if $t>i$. This is not hard to prove, since it is clear that we give $M$ only $(1+\lceil\log i\rceil)S(n)$ space; if $t>i$, then it is impossible to simulate $M_x$, which requires $(1+\lceil\log t\rceil)S(n)$ space, i.e., item (3) in the proof of Theorem \ref{theorem3} will not fit (In fact, we have another approach to define a family of languages $\{L^{'i}_d\}_{i\in\mathbb{N}_3}$ such that $L_d=\bigcup_{i\in\mathbb{N}_3}L^{'i}_d$, given in Appendix \ref{appendixA}).\footnote{In other words, when limited to using only $(1+\lceil\log i\rceil)S(n)$ tape cells, $M$ can only simulate those single-tape deterministic $S(n)$ space-bounded Turing machines for which the number $t$ of tape symbols is less than or equal to $i$.}

Then, we next show the following theorem:
 \begin{theorem}
 \label{theorem4.1}
 It holds that
    $$
    L_d=\bigcup_{i\in\mathbb{N}_3}L_d^i.
    $$
 Furthermore, the following relations hold:
    $$
    L_d^3\subseteq L_d^4\subseteq\cdots\subseteq L_d^i\subseteq L_d^{i+1}\subseteq \cdots.
    $$
 \end{theorem}
 \begin{proof}
  Note first that to accept the language $L_d$, $M$ works within space
$$
  (1+\lceil\log t\rceil)S(n)\quad\text{  for all $t\in\mathbb{N}_3$}.
$$
   This yields 
$$
   L_d = \bigcup_{i\in\mathbb{N}_3} L_d^i.
$$
 
 It is clear that for any fixed $i\in\mathbb{N}_3$,
   $$
   L_d^i\subseteq L_d^{i+1}.
   $$
 Since we can suppose $w\in L_d^i$, i.e., $M$ accepts $w$ within space $(1+\lceil\log i\rceil)S(n)$, $M$ of course can work within space $(1+\lceil\log (i+1)\rceil)S(n)$ to accept $w$, because the $(1+\lceil\log (i+1)\rceil)S(n)$ space is larger than the $(1+\lceil\log i\rceil)S(n)$ space.
 Hence, the proof is completed.
 \end{proof}

Next, for any language $L_d^i$ with $i\in\mathbb{N}_3$, by adapting the proof of Proposition 1.12 in \cite{DK14} (see p. 20, \cite{DK14}), we can show the following specified Tape Compression Theorem:

\begin{theorem}
\label{theorem4.2}
Let $i\in\mathbb{N}_3$ be a fixed positive integer; then 
$$ 
L_d^i\in {\rm NSPACE}[S(n)].
$$
\end{theorem}
\begin{proof}
Note that for any fixed $i\in\mathbb{N}_3$, 
$$
L_d^i\in {\rm NSPACE}[(1+\lceil\log i\rceil)S(n)],
$$
i.e., on input $w$ of length $n$, $M$ scans at most $T(n)$ cells, where
   $$
   T(n)=(1+\lceil\log i\rceil)S(n).
   $$
It is convenient to make, without loss of generality, the assumption about $M$ that the tape cells of $M$ are enumerated by $1,2,3$, etc. We construct a new nondeterministic Turing machine $N$ (using the same number of tapes as $M$) that, on input $w$ of length $n$, simulates the computation of $M(w)$ but works in space $c\cdot T(n)$ with constant $c>0$.

Since $M$ uses the alphabet $\Gamma=\{0,1,\#,\mathrm{b}\}$ of size $4$ and uses $4$ tapes. We suppose $M$ has $r$ states $Q=\{q_1,\cdots,q_r\}$. 

Let $m$ be a positive integer such that $m\geq \frac{1}{c}$. Divide the squares or cells in each work tape of $M$ into groups; each group contains exactly $m$ squares; for example, see Fig. \ref{fig1} when $m=3$. 

\begin{figure}[ht]
\centering
\includegraphics[width=11cm]{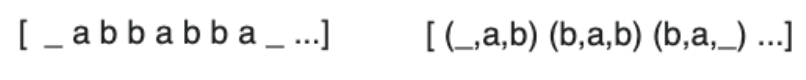}
\caption{\label{fig1} Adjacent groups for $m=3$}
\end{figure}

The machine $N$ uses an alphabet of size $4^m$ (i.e., $|\Gamma|^m$) so that each square of $N$ can simulate a group of squares of $M$. In addition, $N$ has $rm^4$ states, each state represented by $\langle q_i,j_1,\cdots,j_4\rangle$, for some $q_i\in Q$, $1\leq j_1\leq m$, $\cdots$, $1\leq j_4\leq m$. Each state $\langle q_i,j_1,\cdots,j_4\rangle$ encodes the information that the machine $M$ is in state $q_i$ and the local positions of its $4$ tape heads within a group of squares. So, together with the position of its own tape heads, it is easy for $N$ to simulate each move of $M$ using a smaller size of space. For instance, if a tape head of $M$ moves left and its current local position within the group is $j$ and $j>1$, then the corresponding tape head of $N$ does not move, but the state of $M$ is modified so that its current local position within the group becomes $j-1$; if $j=1$, then the corresponding tape head of $N$ moves left and the local position is changed to $m$. By this simulation, $L_d^i=L(M)$ and $L(M)=L(N)$, and $N$ has a space bound $\lceil T(n)/m\rceil\leq c\cdot T(n)$.

Clearly, $N$ use no more than
$$
\lceil T(n)/m\rceil=\left\lceil\frac{(1+\lceil\log i\rceil)S(n)}{m}\right\rceil\leq \lceil c\times (1+\lceil\log i\rceil)S(n)\rceil.
$$
By setting $c=\frac{1}{1+\lceil\log i\rceil}$,
$$
\frac{1+\lceil\log i\rceil}{m}\leq 1
$$
yields
$$
m\geq 1+\lceil\log i\rceil
$$
Thus, let $c=\frac{1}{1+\lceil\log i\rceil}$ (in other words, $m=1+\lceil\log i\rceil$) yield that $N$ scans at most $S(n)$ tape cells for each work-tape of $N$ when simulating $M$ on input $w$ of length $n$. Therefore 
$$
L_d^i\in{\rm NSPACE}[S(n)],
$$
which completes the proof.
\end{proof}

Now, we are ready to show the following important theorem:

\begin{theorem}
\label{theorem4.3}
The language $L_d$ accepted by $M$ is in ${\rm NSPACE}[S(n)]$.
\end{theorem}
\begin{proof}
It is clear that the proof of Theorem \ref{theorem4.2} holds true for arbitrary positive integer $t\in\mathbb{N}_3$. That is to say, for all $t\in\mathbb{N}_3$,
$$
  L_d^t\in{\rm NSPACE}[S(n)].
$$
Thus, Theorem \ref{theorem4.1}, together with Theorem \ref{theorem4.2} and the above observations, yields
$$
L_d\in{\rm NSPACE}[S(n)].
$$
This finishes the proof.
\end{proof}

\vskip 0.3 cm
\begin{remark}
\label{remark4.1}
The above arguments, which confirm our conjecture (i.e., the parameter $t$ is irrelevant to the input length $n$), show that when considering only the growth rates of space functions on the input length of $n$, the parameter $t$, i.e., the number of tape symbols of a deterministic $S(n)$ space-bounded Turing machine, is irrelevant to the language $L_d$'s space complexity.
\end{remark}

\vskip 0.3 cm
\section{Proofs of Theorem \ref{theorem1} and Theorem \ref{theorem2}}
\label{sec:proofs_theorem1_theorem2}
\vskip 0.3 cm

Our goal in this section is to establish Theorem \ref{theorem1}, which states that for $S(n)\geq n$, a space-constructible function, $${\rm DSPACE}[S(n)]\subsetneqq{\rm NSPACE}[S(n)], $$ and to establish Theorem \ref{theorem2}, which gives a negative answer to the LBA question. Indeed, if the above Theorem \ref{theorem3.1} in Section \ref{sec:diagonalization} and Theorem \ref{theorem4.3} in Section \ref{sec:proof_of_ld_in_nspacesn} were not proved, we do not know how to show Theorem \ref{theorem1}. But at this point, with the above two theorems at hand, the proof of Theorem \ref{theorem1} can be made naturally as follows.

\vskip 0.3 cm
\noindent{\it Proof of Theorem \ref{theorem1}.} It immediately follows from Theorem \ref{theorem3.1} and Theorem \ref{theorem4.3}. This finishes the proof. \Q.E.D\\

Indeed, Theorem \ref{theorem2} is a special case of Theorem \ref{theorem1}. To see this, we can set $S(n)=n$. Hence, setting $S(n)=n$ in the proof of Theorem \ref{theorem1}, we can obtain the proof of Theorem \ref{theorem2}.

\vskip 0.3 cm
\noindent{\it Proof of Theorem \ref{theorem2}.}  It clearly follows from Theorem \ref{theorem1} by setting $S(n)=n$. Thus, the proof is completed. \Q.E.D\\

\section{Proof of $\mathit{L}\subsetneqq \mathit{NL}$}
\label{sec:L_NL}
\vskip 0.3 cm

In this section, the main objective is to prove the following theorem with respect to the question of $\mathit{L}$ vs. $\mathit{NL}$, which is a famous and important open question (see e.g., \cite{AB09,Pap94,Sip13}), based on the padding argument like that in Lemma 12.2 in \cite{HU79} (see p. 302, \cite{HU79}). Note that our main computing model in this section is the {\it off-line} deterministic/nondeterministic Turing machine because it has space bounds less than linear, namely $\log n$. Note again that $$\mathit{L}={\rm DSPACE}[\log n] $$ and $$\mathit{NL}={\rm NSPACE}[\log n]. $$

\vskip 0.3 cm
\noindent{\bf Theorem \ref{theorem3}} (restatement)
{\it It holds that:} $$ {\rm DSPACE}[\log n] \subsetneqq {\rm NSPACE}[\log n]. $$ {\it That is,} $$L \subsetneqq NL. $$

\begin{proof}
We show Theorem \ref{theorem3} by contradiction. That is, we assume that $${\rm DSPACE}[\log n] ={\rm NSPACE}[\log n] $$ and obtain a contradiction. So, assuming the result $${\rm DSPACE}[\log n]={\rm NSPACE}[\log n], $$ then we will show by padding argument that
$${\rm DSPACE}[n]={\rm NSPACE}[n], $$which contradicts Theorem \ref{theorem2}.

Now, let $L_1\in{\rm NSPACE}[n]$, i.e., $L_1$ is accepted by a nondeterministic $n$ space-bounded Turing machine $M_1$. Pad the strings in $L_1$ as follows: $$L_2=\left\{ x\$^{(2^{|x|}-|x|)}\,\,|\,\, x\in L_1\right\}, $$ where $\$$ is a new symbol not in the alphabet of $L_1$. Then the padded version $L_2$ is accepted by a nondeterministic Turing machine $M_2$ as follows: first, $M_2$ checks if the input $w$ is of the form $$x\$^{(2^{|x|}-|x|)}. $$ If not then $M_2$ rejects, which requires
$$\aligned
\log \left(|x|+(2^{|x|}-|x|)\right)&=\log 2^{|x|}\\
     &=|x|
\endaligned$$
space; otherwise, $M_2$ marks off $|x|$ cells, then $M_2$ simulates $M_1$ on $x$, accepting if and only if $M_1$ accepts without using more than $|x|$ cells. Note that, the length of input $w$ is
$$\aligned
 |w|&=|x|+(2^{|x|}-|x|)\\
      &=2^{|x|}.
\endaligned$$
Hence, $M_2$ works in space  $$|x|=\log |w|, $$ i.e., $$L_2\in{\rm NSPACE}[\log n]. $$

By the hypothesis that $${\rm NSPACE}[\log n]={\rm DSPACE}[\log n], $$ there is a deterministic $\log n$ space-bounded Turing machine $M_3$ accepting $L_2$. Finally, we construct a deterministic Turing machine $M_4$ accepting the original set $L_1$ within space $n$, which means that $${\rm NSPACE}[n]={\rm DSPACE}[n]. $$

For an input $x$ to $M_4$, $M_4$ marks off $|x|$ cells, which it may do since $n$ is space constructible. $M_4$ has not used more than $|x|$ cells.

Next $M_4$ on input $x$ simulates $M_3$ on $x\$^{(2^{|x|}-|x|)}$. To do this, $M$ must keep track of the head location of $M_3$ on $x\$^{(2^{|x|}-|x|)}$. If the head of $M_3$ is within $x$, $M_4$'s head is at the corresponding point on its input. Whenever the head of $M_3$ moves into the $\$$'s, $M_4$ records the location in a counter. The length of the counter is at most $$\log (2^{|x|}-|x|)\leq\log 2^{|x|}$$ which is less than $|x|$.

If during the simulation, $M_3$ accepts, then $M_4$ accepts. If $M_3$ does not accept, then $M_4$ increases the counter until the counter no longer fits on $|x|$ tape cells. Then $M_4$ halts. Now, if $x$ is in $L_1$, then $x\$^{(2^{|x|}-|x|)}$ is in $L_2$. Thus the counter requires $\log (2^{|x|}-|x|)$ space. It follows that the counter will fit. Thus $x$ is in $L(M_4)$ if and only if $x\$^{(2^{|x|}-|x|)}$ is in $L(M_3)$. Therefore $L(M_4)=L_1$, and $L_1$ is in ${\rm DSPACE}[n]$, which yields $${\rm NSPACE}[n]={\rm DSPACE}[n], $$a contradiction to Theorem \ref{theorem2}. This completes the proof.
\end{proof}

\vskip 0.3 cm
\section{Proof of Theorem \ref{theorem4}}
\label{sec:proofs_theorem4}
\vskip 0.3 cm

Our objective in this section is to establish Theorem \ref{theorem4}, which states that there exists no deterministic $O(\log n)$ space-bounded Turing machine deciding the STCON. 

Before doing the aforementioned work, let us first give some definitions of ``reduction" and ``$\log$-space reduction," which have not appeared in Section \ref{sec:preliminaries}. As noted in \cite{Mon81}, there has been great interest in finding complete problems for various complexity classes defined by Turing machines. This work is important in two aspects. First, to find a complete language for such a class is to show that a single problem represents the complexity of the whole class. Hence, the complexity of the class is better understood. Secondly, to identify a ``natural" problem as being complete for a class is to classify the complexity of this problem; see e.g., \cite{Mon81}.

\begin{definition}[cf. \cite{Mon81}]
 Let $\Sigma$ and $\Delta$ be two alphabets and let $f:\Sigma^*\rightarrow\Delta^*$ be a function. $f$ is $\log$-space computable if there is a deterministic Turing machine with a read-only input tape, an output tape and some work-tapes, which when started with $w\in\Sigma^*$ on its input tape will halt having written $f(w)\in\Delta^*$ on its output tape and having scanned at most $\log|w|$ tape cells on each of its work-tapes.
\end{definition}

\begin{definition}[cf. \cite{Mon81}]
Let $F$ be a class of functions and let $A\subseteq\Sigma^*$, $B\subseteq\Delta^*$ be arbitrary sets. $A$ is $F$-reducible to $B$, denoted by $A\leq_F B$, if there is a function $f\in F$ with $f:\Sigma^*\rightarrow\Delta^*$ such that
$$
\forall w\in\Sigma^*,\,\,\,w\in A\Leftrightarrow f(w)\in B.
$$
We use the terms ``$\log$-space reducible" $(\leq_{\log})$ if $F$ is the class of $\log$-space computable functions.
\end{definition}

Similarly, we have that a question $\mathcal{Q}$ is $\leq_{\log}$--complete for a complexity class $\mathcal{C}$ if the question $\mathcal{Q}$ is in $\mathcal{C}$ and any other questions in $\mathcal{C}$ can be $\log$-space reducible to $\mathcal{Q}$. 

\begin{lemma}[cf. Lemma 4.17, \cite{AB09}]
\label{lemma7.1}
Let $\mathcal{Q}_i$, $i=1,2,3$, be questions.
\begin{enumerate}
  \item [1.]{If $\mathcal{Q}_1\leq_{\log}\mathcal{Q}_2$ and $\mathcal{Q}_2\leq_{\log}\mathcal{Q}_3$, then $\mathcal{Q}_1\leq_{\log}\mathcal{Q}_3$.}
  \item [2.]{If $\mathcal{Q}_1\leq_{\log}\mathcal{Q}_2$ and $\mathcal{Q}_2\in L$, then $\mathcal{Q}_1\in L$.\Q.E.D}
\end{enumerate}
\end{lemma}

At this point, with the notions given above and by Theorem \ref{theorem3} at hand, the proof of Theorem \ref{theorem4} can be made naturally by a contradiction to Theorem \ref{theorem3}.

\vskip 0.3 cm
\noindent{\it Proof of Theorem \ref{theorem4}.}  Since STCON is $\leq_{\log}$--complete for the complexity class $\mathit{NL}$ (see e.g., \cite{AB09,Pap94,Sip13}), from which we know that 
$$
\mathit{L}=\mathit{NL}
$$
if and only if 
$$
{\rm STCON} \in \mathit{L},\quad\text{(by Lemma \ref{lemma7.1})}
$$ 
i.e., if and only if there is a deterministic algorithm deciding STCON in space $O(\log n)$.

Now, if STCON has deterministic algorithms with space complexity $O(\log n)$, we have
$$
\mathit{L}=\mathit{NL},
$$
a contradiction to Theorem \ref{theorem3}. This completes the proof. \Q.E.D

\vskip 0.3 cm
\section{Concluding Remarks}
\label{sec:conclusion}
\vskip 0.3 cm

To summarize, we have shown that the class of languages accepted by nondeterministic linear-bounded automata (i.e., the context-sensitive languages) is not equal to the class of languages accepted by deterministic linear-bounded automata. Thus, we resolve the LBA question, which is a famous and longstanding open question in {\it automata theory}.

In fact, we have shown a more general result, i.e., for any space-constructible function $S(n)\geq n$, $${\rm DSPACE}[S(n)]\subsetneqq{\rm NSPACE}[S(n)]. $$ We achieved this by enumerating all deterministic $S(n)$ space-bounded Turing machines, then diagonalizing against them by a universal nondeterministic Turing machine $M$. After that, we use novel and interesting methods to show that the language $L_d$ accepted by the simulating machine $M$ is in fact in ${\rm NSPACE}[S(n)]$.

Next, we study the relationship between ${\rm DSPACE}[\log n]$ and ${\rm NSPACE}[\log n]$ and show by padding argument that $${\rm DSPACE}[\log n]\subsetneqq{\rm NSPACE}[\log n], $$i.e.,$$\mathit{L}\subsetneqq \mathit{NL}, $$which also resolves a famous open question in complexity theory. It is clear that the result of$$\mathit{L}\subsetneqq \mathit{P}$$ follows obviously from $$\mathit{L}\subsetneqq \mathit{NL}.$$

As a special by-product of our result, the STCON has been resolved as well. That is, no deterministic $\log n$ space-bounded Turing machine can decide the $st$-connectivity of a directed graph, where $n$ is the number of vertices in that directed graph. This is because the STCON is $NL$-complete under the log-space reduction.

Finally, we should say that there are many important questions we did not touch on in this paper; see e.g., \cite{A1}. Moreover, we have only shown $${\rm DSPACE}[S(n)]\subsetneqq{\rm NSPACE}[S(n)]$$ for $S(n)\geq n$ and $S(n)=\log n$, and leaving open the case for $$\log n<S(n)<n. $$
In addition, it was shown in \cite{RCH91} that we could separate nondeterministic space from deterministic space by showing a nondeterministically fully space-constructible function below $\log n$. In particular, the work \cite{Gef91} pointed out that if $\log\log n$ were fully space constructible by a nondeterministic Turing machine, then we would have $${\rm DSPACE}[S(n)]\neq{\rm NSPACE}[S(n)]$$ for $\log \log n\leq S(n)< \log n$.

\subsection*{Acknowledgements}
Sincere thanks go to the anonymous reviewers whose useful comments corrected the author's incorrect perception of earlier versions of this manuscript.

\bibliographystyle{ACM-Reference-Format}

\appendix
\section{Alternative Construction of $\{L^{'i}_d\}_{i\in\mathbb{N}_3}$}
\label{appendixA}

Note that the language $L_d^{'i}$ is accepted by $M$ while running in $$(1+\lceil\log i\rceil)S(n)$$ space. This means that $M$ simulates those single-tape deterministic Turing machines with $t=|\Gamma|$ for which $t\leq i$, since the simulation takes $$(1+\lceil\log t\rceil)S(n)$$ space and there are $(1+\lceil\log i\rceil)S(n)$ cells of space in total. Thus, for a fixed $i\in\mathbb{N}_3$, the language $L_d^{'i}$ can be constructed as follows:

Since all single-tape deterministic Turing machines are enumerable, let $M$ be a four-tape NTM that operates as follows on an input string $x$ of length $n$.
\begin{enumerate}
\item{ By using 
$$
O(\log |x|)
$$
space, $M$ decodes the single-tape deterministic Turing machine encoded by $x$. If $x$ is not the encoding of some single-tape deterministic Turing machine, then GOTO $6$. Otherwise, $M$ decodes the deterministic Turing machine encoded by $x$ and determines $t$, the number of tape symbols used by this deterministic Turing machine, and $s$, its number of states. The third tape of $M$ can be used as ``scratch" memory to calculate $t$. }
\item {\color{red}If $t\leq i$ then GOTO (3); else reject the input.}\label{item2}
\item {$M$ marks off $(1+\lceil\log t\rceil)S(n)$ cells on each work tape. After doing so, if any tape head of $M$ attempts to move off the marked cells, $M$ halts without accepting.}
\item {Let $M_l$ be the deterministic Turing machine encoded by $x$.  $M$ then lays off on its second tape $S(n)$ blocks of $\lceil\log t\rceil$ cells each, the blocks being separated by single cells holding a marker $\#$, i.e., there are $(1+\lceil\log t\rceil)S(n)$ cells in all. Each tape symbol occurring in a cell of $M_l$'s tape will be encoded as a binary number in the corresponding block of the second tape of $M$. Initially, $M$ places its input, in binary-coded form, in the blocks of tape $2$, filling the unused blocks with the code for the blank.}
\item {On tape $3$, $M$ sets up a block of $\lceil\log s\rceil+\lceil\log S(n)\rceil+\lceil\log t\rceil S(n)$ cells, initialized to all $0$'s, provided again that this number of cells does not exceed $(1+\lceil\log t\rceil)S(n)$, which is possible, because $$\aligned
            \frac{\lceil\log s\rceil+\lceil\log S(n)\rceil+\lceil\log t\rceil S(n)}{(1+\lceil\log t\rceil S(n)}\rightarrow& \quad\frac{\lceil\log t\rceil}{1+\lceil\log t\rceil}\\
            <&\quad 1
             \endaligned$$ as $$ n\rightarrow +\infty
            $$
            and since there are arbitrarily long binary strings representing the same Turing machine $M_l$.
            Tape $3$ is used as a counter to count up to $sS(n)t^{S(n)}$.
            }
\item {$M$ simulates $M_l$, using tape $1$ (its input tape) to determine the moves of $M_l$ and using tape $2$ to simulate the tape of $M_l$. The moves of $M_l$ are counted in binary in the block of tape $3$, and tape $4$ is used to hold the state of $M_l$. If $M_l$ accepts, then $M$ halts without accepting. $M$ accepts if $M_l$ halts without accepting. Moreover, $M$ will halt without accepting if the simulation by $M$ attempts to use more than the allotted cells on tape $2$, or if the counter on tape $3$ overflows, i.e., the number of moves made by $M_l$ exceeds $sS(n)t^{S(n)}$.}\label{item5}
\item{Since $x$ is not the encoding of some single-tape deterministic Turing machine, $M$ marks off 
$$
S(n)
$$
cells on each work tape. After doing so, if any tape head of $M$ attempts to move off the marked cells, $M$ halts without accepting. Then, on tape $3$, $M$ sets up a block of 
$$
S(n)
$$
cells, initialized to all $0$'s. Tape $3$ is used as a counter to count up to 
$$
2^{S(n)}.
$$

By using its nondeterministic choices, $M$ moves according to the path described by $x$. The moves of $M$ are counted in binary in the block of tape $3$. $M$ rejects if the number of moves made by $M$ exceeds 
$$
2^{S(n)}
$$
or $M$ reaches a reject state before the counter of tape $3$ reaches 
$$
2^{S(n)}.
$$
Otherwise, $M$ accepts. }
\end{enumerate}

Compared with the proof of Theorem \ref{theorem3.1}, the construction of $M$ differs only in Item \ref{item2}.

Nevertheless, since all single-tape deterministic Turing machines are simulated by $M$ as $i$ runs through all numbers in $\mathbb{N}_3$, we certainly have $$L_d=\bigcup_{i\in\mathbb{N}_3}L_d^{'i}. $$

\end{document}